\documentstyle[prb,aps]{revtex}
\newcommand{\dbs}{\renewcommand{\baselinestretch}{1.5}
\large\normalsize}
\dbs
\begin{document}
\title{
Slip Energy Barriers in Aluminum and Implications 
for Ductile versus Brittle Behavior}
\author{
Yuemin Sun$^{(a),\dagger}$
and Efthimios Kaxiras$^{(a,b),\star}$\\
}
\address{
$^{(a)}$ Division of Applied Sciences,
Harvard University, Cambridge MA 02138\\
$^{(b)}$ Department of Physics,
Harvard University, Cambridge MA 02138\\
}
\maketitle
\begin{abstract}
We conisder the brittle versus ductile behavior of aluminum
in the framework of the Peierls-model analysis of  
dislocation emission from a crack tip.  
To this end, 
we perform first-principles quantum mechanical calculations for the 
unstable stacking energy $\gamma_{us}$ of aluminum along the 
Shockley partial slip route.  Our calculations are based on 
density functional theory and the local 
density approximation and 
include full atomic and volume relaxation.
We find that in aluminum $\gamma_{us} = 0.224$ J/m$^2$.  
Within the Peierls-model analysis, this value would predict a 
brittle solid which poses an interesting problem since 
aluminum is typically considered ductile.  The resolution may be 
given by one of three possibilites: (a) Aluminum is indeed brittle 
at zero temperature, and becomes ductile at a finite temperature due to 
motion of pre-existing dislocations which relax the stress 
concentration at the crack tip. 
(b) Dislocation emission at the crack tip is itself a thermally 
activated process.
(c) Aluminum is actually ductile at all temperatures and the 
theoretical model employed
needs to be significantly improved in order to resolve the apparent 
contradiction.
\end{abstract}


\section{Introduction}

Understanding the ductile versus brittle (D/B) response of materials is both 
scientifically challenging and technologically important.
The D/B response of most metals is usually established by
experimental methods.  A theoretical framework that can 
describe quantitatively D/B behavior has been pursued for over two decades,
beginning with the seminal work of Rice and Thomson (1974).
When dealing with complicated high-performance materials, like intermetallic 
compounds, silicides, etc. the ability to predict the D/B
behavior from theoretical considerations becomes even more 
important, since experimental measurements are not available, or are 
difficult and time consuming.
So far, approximate estimates are available
for certain systems, based on atomistic simulations that 
employ simple interatomic potentials [Cheung (1990); Sun, Rice
and Truskinovsky (1991); Beltz and Rice (1992); Sun, Beltz and Rice (1993)].
Using a Peierls type of analysis [Peierls (1940)], Rice and coworkers
[Rice (1992); Rice, Beltz and Sun (1992); Sun, Beltz and Rice (1993);
Rice and Beltz (1994);
Sun and Beltz (1995)] have recently developed simple criteria to characterize
the D/B behavior.  These criteria use few key parameters related 
to the properties of the solid,
namely the unstable stacking energy, the surface energy, 
the shear modulus and the Burgers vector.  Zhou, Carlsson and 
Thomson (1993, 1994) have also developed similar criteria
that involve these parameters, 
using atomistic studies of model systems.
Of these parameters, the one that is 
not accessible experimentally is the unstable stacking 
energy $\gamma_{us}$, identified by Rice (1992) to be the 
quantity that controls the emission of straight dislocations from the
crack tip under shear loading. 
The value of $\gamma_{us}$ is the lowest energy barrier 
that needs to be surmounted when one half of a crystal 
slides over the other half in going from one ideal configuration 
to another equivalent one (the lowest barrier may actually 
occur between an ideal and a metastable configuration, 
corresponding to nucleation of partial dislocations). 
The importance of this quantity within the Peierls model
makes it desirable to obtain 
as accurate estimates as possible for $\gamma_{us}$ 
in various materials. 
Kaxiras and Duesbery
(1993) have used first-principles quantum mechanical 
calculations in the context of density 
functional theory to obtain the value of the unstable stacking 
energy for silicon, a prototypical covalent material. 
Here we performed similar calculations for aluminum, a representative
simple metal, and analyze the implications of the results
for the D/B behavior of aluminum.
Related work on the theoretical strength of aluminum using 
first-principles quantum mechanical calculations has been 
reported by Paxton et al. (1991).

The metallic nature of Al and the relatively small energy 
cost for the slip (compared, for example, to Si) require more attention 
to computational details.
Before embarking on the calculation 
of $\gamma_{us}$, we performed several tests to determine 
the limitations of our calculations.  These tests 
are described in Section II.  Section III discusses our results 
for $\gamma_{us}$. 
Section IV concludes with some discussion of the 
implications of our results for the D/B behavior of
Al in the context of current theories.

\section{First-principles calculations for aluminum}

\subsection{Bulk properties}

Our first-principles calculations are based on
density functional theory [Hohenberg and Kohn (1964)]
in the local density approximation [Kohn and Sham (1965)] (in the 
following referred to as DFT/LDA).
We employ the expression for the exchange and correlation 
functional proposed by Perdew and Zunger (1981), and 
a norm-conserving non-local pseudopotential 
from Bachelet, Haman and Schl\"{u}ter (1982) to represent 
the atomic core and eliminate the core electrons of Al.  
A plane wave basis is used to expand 
the wave functions of the Kohn-Sham orbitals.  
Since the physical quantities of interest involve obtaining 
small energy differences by subtracting large numbers, particular 
care must be taken to assess the uncertainty in these numbers.  There
are two sources of errors:  The first has to do with computational choices, 
such as limiting the plane wave 
basis to sets with kinetic energy up to a maximum value, and 
approximating integrals over the Brillouin Zone (BZ) 
by sums over finite sets of reciprocal space points,
referred to as k-points; the other source of error has to do  
with inherent limitations of the formalism we employ.
We attempted to minimize the first type of error 
by variationally expanding the plane wave basis 
and by enlarging the sets of k-points used in reciprocal 
space integrations.  The theoretical results are
reasonably well converged with respect to these computational choices. 
In order to provide estimates 
of how much the converged results differ 
from true physical values, we compare to 
the values of quantities that can be measured experimentally,
specifically the energy of the intrinsic stacking 
fault in Al.  
Any residual difference is the error 
inherent in the calculations due to fundamental limitations
of the formalism.  We discuss in the concluding section
how the present results may be used to extract 
useful insight despite their limitations.

As a first test we have calculated the equilibrium lattice 
constant and the elastic properties of fcc bulk Al.  
For the equilibrium lattice constant we use a cutoff in 
the kinetic energy of plane waves equal to 12 Ry (corresponding 
to 70 plane waves per atom) and a uniform grid of k-points in 
the BZ produced by dividing each of the three
primitive vectors in reciprocal space in intervals
of equal size [Monkhorst and Pack (1976)].
We use the notation $b_i/\delta k_i = n_i$
to denote the number of divisions in the primitive 
reciprocal space vectors $(i=1,2,3)$, where $b_i$ is the magnitude of 
a vector and $\delta k_i$ is the interval corresponding 
to a certain choice of $n_i$.   
In the case of bulk fcc Al, $b_1 = b_2 = b_3$
and therefore we take $n_1 = n_2 = n_3 = 16$.
These cutoffs are adequate for well 
converged calculations of bulk properties.  
In order to obtain the equilibrium lattice constant from the 
computed energy as a function of lattice constant, we fit 
to the universal 
binding energy relation proposed by Rose et al. (1984),
\begin{equation}
E(a^*) = E_0 + E_c - E_c (1+a^*) \exp(-a^*)
\end{equation}
where $E_0, E_c$ correspond to the 
minimum energy and the cohesive energy respectively,
and the reduced lattice 
constant $a^*$ is given in terms of the actual lattice constant $a$ 
as $a^* = (a - a_0) / l$ 
with $a_0$ the equlibrium lattice constant and $l$ 
a parameter setting the length scale.  In this expression,
$E_0, a_0, l$ are viewed as fitting parameters and $E_c$ is
taken from experiment $E_c = 3.39$ eV per atom.  
To obtain the values of
the parameters $E_0, a_0, l$ we calculated the energy at 
25 points between $a$ = 3.6 and 6.4 \AA.  We
find $a_0 = 3.95$ \AA.
The experimental lattice constant of Al at room temperature
is 4.05 \AA.  In order to compare our theoretical value of
the equlibrium lattice constant (which of course is calculated 
at zero temperature) to the experimental one at room temperature, 
we use the experimental 
thermal expansion coefficient $\alpha = 2.36 \times 10^{-5} K^{-1}$
(Pearson (1958)), 
to extrapolate between zero and room temperature.
This gives a theoretical estimate for the lattice constant 
at room temperature of 4.02 \AA, which 
is in excellent agreement with experiment.
We also calculated the bulk modulus $B$ from the second 
derivative of the energy with respect to volume,
evaluated at the point where the first derivative 
vanishes.  
We obtain $B = 84.8$  GPa compared to the experimental 
result of 76.93 GPa.  This difference of 10 \% is typical
of DFT/LDA calculations. 

In addition to $a_0$ and $B$, we have calculated the elastic 
constants $C_{11} - C_{12}$ and $C_{44}$, which enter in the 
expression of the shear modulus
$\mu = (C_{11} - C_{12} + 3 C_{44})/5$.  
The elastic constants were obtained by using the stress-strain 
relations, and inducing appropriate distortions of the 
unit cell.  The amount of the distortion was large enough to 
produce energy differences that can be calculated accurately,
yet small enough so that quadratic fits to the energy
are appropriate.  Typical distortions were in the range of 5 - 10 \%.
We have performed 
these calculations with a higher density 
of sampling points in the BZ.  This was deemed necessary because 
the calculation of elastic constants involves rather small 
energy differences.  We find that their values are converged 
for $n_1 = n_2 = n_3 = 20$.  
The values of the elastic constants and the shear modulus 
are sensitive to the lattice constant of the crystal.
Accordingly, we have performed the calculation at two 
different lattice constants, the theoretical one at zero temperature
($a_0 = 3.95$ \AA), and the experimental one at room temperature
($a_0' = 4.05$ \AA).  The results are tabulated in
Table I.  The agreement with experiment is poor at the 
the theoretical lattice constant at zero temperature, but becomes 
reasonable at the room-temperature experimental lattice constant.
Our results compare favorably with previous calculations
performed at the experimental lattice constant $a_0'$
by Mehl and Boyer (1991).

\subsection{Intrinsic stacking fault energy}

As a final test of the reliability of our approach
we have calculated the value of the intrinsic stacking 
fault $\gamma_{isf}$ in aluminum, 
a number that can be determined experimentally.
An additional advantage of 
performing this calculation is that 
the technical aspects 
are identical to the calculation of the 
unstable stacking energy.  Specifically, 
both the intrinsic stacking fault and the unstable stacking 
energy can be obtained by considering a slab and shearing 
it in a periodic fashion by a certain distance.  This is ilustrated
in Fig. 1: in 1(a), a top view of the ABCABC stacking 
of layers in the fcc lattice is displayed with high symmetry 
directions identified; in 1(b), a side view is shown, 
with the slip plane identified.  The periodic slab consists 
of an integer multiple of ABC layers in the [111] direction.
In our clalculations we have used slabs with two and three 
periods (i.e. consisting of 6 and 9 layers in the [111] direction),
to check convergence with respect to slab size.  
For slip of $a_0/\sqrt{6}$ in the 
$[12\bar{1}]$ direction, one obtains the intrinsic stacking fault 
configuration, i.e. a structure that involves the 
stacking ABCBCABC, with the stacking fault between the 
third and fourth layers in the sequence.  The unstable stacking 
energy corresponds to a configuration that is sheared partly
from the ideal configuration to the intrinsic stacking fault
one by an as yet unspecified amount.  

For both the intrinsic stacking fault and the 
unstable stacking energy, we performed the calculations at the 
theoretically determined lattice constant.  This is an important 
detail that deserves justification:  the structure of 
defects, such as vacancies, interstitials, stacking faults, 
grain boundaries, etc. involves relaxation of the atomic
coordinates to a fully optimized geometry
in which the calculated forces on the ions are vanishingly small.
In order for the relaxed configuration to make physical sense,
one has to keep the crystal far from the defect at its 
equilibrium lattice constant.  In a supercell calculation, 
the boundaries of the 
unit cell represent this ``far from the defect'' region.  
This ensures that any atomic relaxation 
in the neighborhood of the defect is the result of the 
presence of the defect, rather than externally imposed strain. 
Thus, defect calculations,
including the stacking fault ones, must be performed at the 
theoretical lattice constant. 

In the slab calculations, the three lattice vectors 
are no longer equivalent.  The two planar vectors are actually
identical to two primitive fcc lattice vectors,
for instance along the [110] and [101] directions,
as indicated in Fig. 1(a).  The third lattice vector is in 
the [111] direction, and is a multiple of the repeat 
distance between A layers in that direction.  
Because of these differences, care must be taken to 
perform the calculations at the same level of convergence
as the bulk calculation, so that the energy of the 
intrinsic stacking fault and the unstable stacking can be compared 
to that of the bulk.  Specifically, the number of intervals
in each reciprocal space direction must be proportional to the length of 
the reciprocal lattice vector, or equivalently, inversely
proportional to the length of the real-space repeat vector. 
Moreover, the angles between the various vectors must 
also be taken into account, so that the volume density 
of reciprocal points in the calculations remains 
approximately the same.  Since in the fcc lattice the 
primitive vectors form 60$^0$ angles, whereas in the 
slab calculation with one vector along the [111] direction
two of the angles are 90$^0$, an extra factor of $\sin(60^0)$
must be included in figuring out the ratios of divisions 
along the reciprocal lattice directions.  With these 
considerations, we find that the following relations
must hold:
$n_1 = n_2 = 3 \sqrt{2} n_3 $ and $n_1 = n_2 = 9/\sqrt{2} n_3$
for the 6-layer and 9-layer slabs respectively,
where the repeat vector in the [111] direction is identified
with the index 3.  These relationships cannot be satisfied 
exactly for integer divisions of the reciprocal lattice 
vectors.  Instead, we have used the relations
$n_1 = n_2 = 4 n_3 $ and $n_1 = n_2 = 6 n_3$
which satisfy the desired ratios to a good approximation.
Convergence tests were performed for both the number of 
divisions along each reciprocal space direction, as 
well as the number of plane waves per atom. 
The results of these tests 
are shown in Fig. 2.  

Although the results for the 6-layer and 9-layer supercells
are reasonably close, indicating adequate convergence 
with respect to supercell size, we have performed an additional test
to establish how reliable these numbers are.      
For this test we used the anisotropic next-nearest-neighbor Ising
(ANNNI) model to obtain the energy of the intrinsic stacking fault. 
Details of this approach can be found in the work
of Denteneer and Soler (1991a, 1991b).   
The calculation of the intrinsic 
stacking fault energy is based on assuming coupling constants between 
different layers, and obtaining the values of 
the coupling constants by comparing the energies of various
periodic stackings, such as ABC (corresponding to the fcc lattice),
AB (corresponding to the hcp lattice) and ABCB.  
Keeping only the lowest order terms, the first two coupling 
constants $J_1, J_2$ are given by
\begin{eqnarray}
J_1 = \frac{1}{4} E(AB) - \frac{1}{6} E(ABC) \\
J_2 = \frac{1}{4} E(ABCB) - \frac{1}{6} E(ABC) - \frac{1}{2} J_1 
\end{eqnarray}
and the intrinsic stacking fault is given by
\begin{equation}
\gamma_{isf} = 4 (J_1 + J_2) / A
\end{equation}
where $A = \sqrt{3} a_0^2/4$ 
is the area per unit cell on the plane of the fault. 
The advantage of the approach is that very small unit cells
can be used to extract the values of the coupling constants,
allowing for more extensive convergence tests.
The results for $\gamma_{isf}$ obtained from the ANNNI model 
are included in Fig. 2 for comparison to the supercell calculations.
Taken together, these results indicate that the value of the 
instrinsic stacking fault is reasonably well converged 
with a basis of 70 plane waves per atom (corresponding to 
a kinetic energy cutoff of 12 Ry), and for a density of 
BZ sampling points corresponding to 16 divisions along the 
direction of primitive in-plane vectors.
The value obtained from these calculations is 
\begin{equation}
\gamma_{isf} = 0.165 \pm 0.015  \; \mbox{J/m$^2$}
\end{equation}
The error bar was estimated 
by assuming that the contributions
from (a) the size of the plane-wave basis, (b) the density 
of BZ sampling points, and (c) the size of the supercell,
are independent, as experience with
similar calculations and Fig. 2 indicate.
The total error is then obtained 
as the square root of the sum of squares of the three contributions.

The value of $\gamma_{isf}$ that we obtained 
is in excellent agreement with the 
result of Wright et al. (1992) 0.161 J/m$^2$, who used a similar method 
in their calculations (the plane-wave pseudopotential approach).  
Our value for $\gamma_{isf}$ is
higher than the result of Denteneer and Soler (1991a),
0.126 J/m$^2$, which was  
obtained with a different computational method (the Augmented Plane 
Wave (APW) approach). 
Experimental measurements range from a low of 0.110 J/m$^2$ to a high of 
0.280 J/m$^2$, with the most recent result at 0.150 J/m$^2$
(by Mills and Sadelmann (1989) - see Wright et al. 
(1992) for additional information).   
Our calculated value is also in excellent agreement 
with the experimental value 0.166 J/m$^2$
quoted by Hirth and Lothe (1982).

\section{Unstable stacking fault}

The value of the unstable stacking energy 
$\gamma_{us}$ was obtained by shearing half of the infinite 
crystal over the other half and finding the lowest 
energy barrier that needs to be overcome in order 
to bring the crystal from one ideal configuration
to another equivalent one. 
The path along which the energy barrier is lowest is in one of 
the equivalent <211> crystallographic directions,
for instance the $[12\bar{1}]$ direction, shown in Fig. 1.
Slip along this direction by 
\begin{equation}
b_p = a_0/\sqrt{6}
\end{equation}
leads to the
instrinsic stacking fault configuration, which corresponds
to a metastable configuration.
The unstable stacking energy configuration corresponds 
to the saddle point along the path from the equilibrium to the 
metastable configuration, which should occur near,
but not necessarily at, the midpoint.
To determine the position of the saddle point we calculated 
the energy for several displacements along the slip path.
We refer to these energies as the generalized stacking fault
energy $\gamma_{gsf}$.  The results are shown in Fig. 3.
Our calculations indicate that the saddle point configuration
occurs at $0.62 b_p$.
The value of the energy at the 
saddle point configuration, before 
any relaxation is taken into account, is
\begin{equation}
\gamma_{us}^{(u)} = 0.249 \; \mbox{J/m$^2$}.
\end{equation}

When both atomic 
relaxation and volume relaxation are taken into account,
this value drops by 0.025 J/m$^2$.
Within the Peierls model framework, 
volume relaxation is meaningful only in the [111] direction.
Our estimate 
of the relaxed unstable stacking energy in aluminum is 
\begin{equation}
\gamma_{us}^{(r)} = 0.224  \; \mbox{J/m$^2$} .
\end{equation} 
Both atomic relaxation and volume relaxation
were calculated using the 
theoretically determined value for the in-plane lattice constant,
i.e. $a_0 = 3.95$ \AA, so that spurious contributions
from strain in the lattice are avoided, as explained in the 
previous section.
Since the unstable stacking energy was obtained by exactly the same 
computational parameters as the intrinsic stacking fault
energy, we expect that the same error bars as determined 
in the previous section will apply.  

It is useful to express
the above results for the unrelaxed and relaxed 
values of the unstable stacking energy
in terms of the dimensionless quantities 
defined by Sun, Beltz and Rice (1993), that provide 
estimates of the importance of tension-shear coupling.
These quantities are defined as: 
\begin{equation}
q = \frac{\gamma_{us}^{(u)}}{2 \gamma_s}, \; 
p = \frac{\Delta_{\theta}^{*}}{L} , 
\end{equation} 
where $\gamma_s$ is the energy per unit area of the surface 
exposed during decohesion,  
$\Delta_{\theta}^{*}$ is the value of the opening 
displacement when atomic relaxation is included and 
corresponding to tensile stress $\sigma = 0$
at the unstable stacking configuration, 
and $L$ is a phenomenological length scale for tension.
The values of the various quantities in the above equations 
as obtained from the present 
calculations and from previous work by Sun, Beltz and Rice (1993)
using the embedded atom method (EAM), are given in Table II.
In this comparison, we have used the value of
$\gamma_s = 1.10$ J/m$^2$, from the work of 
Ferrante and Smith (1979), which was obtained using DFT-LDA
calculations.  
The comparison of the two sets of results in Table II,
one from the present first-principles calculations,
the other from the empirical EAM calculations, reveals that
while the bare 
quantities $\gamma_s, \gamma_{us}^{(u)}, \gamma_{us}^{(r)}$
differ by factors of 2 to 3 in the two calculations, the
dimensionless scaled quantities $p, q, L$ are actually rather close.
Apparently, the underestimates of the bare quantities 
by large factors in the EAM calculations cancel out
when the scaled quantities $p$ and $q$ are computed, giving reasonable 
estimates of the shear to tension coupling. 

\section{Implications for brittle-ductile behavior}

Significant progress has been made recently
in developing criteria for the D/B behavior 
by the theoretical analysis of Xu, Argon and Ortiz (1995,1996a,1996b)
using the boundary integral method, and the studies 
by Zhou, Carlsson and Thomson (1993,1994), using a Green's function
approach and model atomistic systems.  
The involved nature of these studies 
reflects the inherent difficulty in capturing a very 
complex dynamical phenomenon such as brittleness 
or ductility, with a few parameters.  
In the spirit of retaining a simple criterion
for the D/B  behavior, 
it is worthwhile to examine the implications of the 
present results for aluminum. 
We will consider two different contexts: 
The first is based on the criteria developed 
by Rice and coworkers 
[Rice (1992); Rice, Beltz and Sun (1992); Sun, Beltz and Rice (1993);
Rice and Beltz (1994);
Sun and Beltz (1995)], the second derives from the atomistic 
studies of model systems by Zhou et al. (1993,1994).

In the context of the Peierls type analysis
of Rice and coworkers, one can obtain the critical loading
for dislocation emission at a crack tip $G_d$, which depends
on the value of $\gamma_{us}$ and the external loading
(assumed here to be mode I), 
as well as the 
geometry of the dislocation to be emitted.
In the case of aluminum, the relevant dislocation is a
Shockley partial, and the parameters entering in the 
geometry are the tilt angle $\theta$ between the 
the slip plane and the extension of the 
crack, and the angle $\phi$ between the direction 
in which the dislocation is emitted and its Burgers vector.
These features are illustrated in Fig. 4.  
For the case of tension-shear coupling, 
the value of $G_d$ (typically given in terms of $\gamma_s$,
i.e. the ratio $G_d / 2 \gamma_s$), is obtained 
by solving numerically 
a pair of coupled integral equations as described 
in detail by Sun, Beltz and Rice (1993).  
We have performed such calculations for the slip 
route described earlier, 
for several crack geometries described by the 
three vectors $\vec{a}_1, \vec{a}_2, \vec{a}_3$,
pointing along the crack propagation direction, normal to the crack 
plane and along the crack line, respectively (see Fig. 4).
In these calculations we take the surface energy $\gamma_s$
to be the same for the 
(111) and (001) surfaces (which is true to a good approximation),
and we use the experimental values for the 
elastic constants from Hirth and Lothe (1982).

The results of our calculations are shown in Table III.  
The geometry labeled  
A, with $\{ \vec{a}_1, \vec{a}_2, \vec{a}_3 \} = 
\{ [\bar{1}10], [001], [110] \}$, i.e. with a crack 
on the (001) plane, is interesting in that it produces a ratio 
$G_d / 2 \gamma_s > 1$.
The easiest slip system for this geometry 
is on the $(1\bar{1}1)$ plane, along the $[121]$
direction, with corresponding angles $\theta = 54.7^0$ and
$\phi = 60^0$.  We find that for this geometry,
$G_d / 2 \gamma_s = 1.74$,  a value which
indicates that dislocation emission is energetically
unfavorable compared to crack propagation by cleavage,
since according to the Griffith criterion [Griffith (1920)] 
the critical loading for cleavage is $G_c = 2 \gamma_s$.  
This result then implies that all cracks on (001) planes are
brittle and that (001) planes in aluminum are intrinsically cleavable.
In light of experimental 
evidence that suggests aluminum to be ductile, this 
is a surprising result.  We return to this point below.

A different way to link the present results to the 
intrinsic brittleness or ductility of aluminum, is through 
comparison to recent atomistic calculations on model systems.
The work of Zhou et al. (1993,1994) has investigated the conditions
for ductile vs. brittle behavior in a model material,
which consisted of a two-dimensional 
solid of atoms interacting through a variety 
of empirical potentials.
These authors find that the ratio $\gamma_{us}^{(r)} / \mu b$ provides 
a useful means for characterizing the D/B
behavior, with the value 0.015 separating the two regimes,
and ductile behavior corresponding to smaller values of the ratio.
From the results of the present study, we find that
\begin{equation}
\frac{\gamma_{us}^{(r)}}{\mu b} = 0.0287
\end{equation}
when we use the value of $\mu$ calculated at the 
theoretical lattice constant of 3.95 \AA.
Our value for the ratio that characterizes D/B 
behavior is larger than the criterion of   
Zhou et al. (1993,1994) by almost a factor of 2, 
again implying a brittle behavior for 
aluminum within this theoretical model.

One may question the ability of the present 
calculations to obtain accurate estimates of the fundamental 
quantities entering the D/B criteria.
In defense of the accuracy of the present calculations 
(putting aside the careful comparison of theoretical 
results to available experimental numbers discussed 
in Section II), we invoke the following argument:
Suppose that 
experimental numbers were used exclusively for the 
values of key quantities.  A strict lower bound for
the value of $\gamma_{us}$ is 
the value of $\gamma_{isf}$,
since the unstable stacking energy
cannot be lower than the energy of the instrinsic stacking
fault.  Using the value of $\gamma_{isf}$ 
as an approximation to $\gamma_{us}$
and the experimental values for 
$\mu = 26.5$ GPa and $b = a_0/\sqrt{6}$, $a_0 = 4.05$ \AA, 
one would obtain a ratio of
$\gamma_{us} / \mu b$ 
in the range of 0.025 to 0.064
(from the experimental values for $\gamma_{isf}$
which range from 0.11 to 0.28 J/m$^2$).  This 
is a lower bound (since $\gamma_{isf}$ is a lower bound for 
$\gamma_{us}$) and 
is still 
much higher than the value of 0.015 proposed by Zhou 
et al. (1993,1994) as separating brittle from ductile behavior.
In fact, the lowest value of this estimate is reasonably 
close to the result obtained from our first-principles 
calculations.  This argument leads us to suggest that 
the surprising result obtained here, namely
that aluminum is predicted 
to be brittle, is not due to limitations of the first-principles 
calculations.

We also wish to point out that surface energy
terms associated with surface creation during dislocation emission 
and lattice trapping were not taken into account in the
Peierls type analysis discussed above.  Both 
of these effects would tend to {\em increase} the value 
of the ratio $G_d / 2 \gamma_s$ relative to what has 
been reported here [see for example, Xu, Argon and Ortiz (1995); 
Juan, Kaxiras and Sun (1996)], 
titling the balance toward more brittle
behavior.  On the other hand, the atomistic simulations
of Zhou et al. (1993,1994) which do take these effects into account,
they nevertheless provide a picture consistent with the Peierls analysis.  

These results pose an interesting puzzle.  To our knowledge,
there is no experimental indication of brittle behavior 
in aluminum at finite temperature.  The resolution of 
the puzzle may be provided by three different possibilities:
(a) The theoretical framework invoked to discuss  D/B
behavior applies to the {\em intrinsic} behavior of 
a pure material; the behavior of a material with 
high density of pre-existing dislocations is not captured 
by this framework.  Thus, aluminum may be ductile due to 
motion of pre-existing dislocations.  
Beltz, Rice, Shih and Xia (1996) recently addressed the issue
of crack growth in the presence of a large number 
of pre-existing dislocations.  Their analysis could provide  
important insight to the problem discussed here.
(b) Since, strictly speaking, 
the theoretical analysis presented above applies 
to zero temperature, aluminum may indeed be brittle at zero 
temperature and its ductility is due to thermally activated 
dislocation emission.
Rice and Beltz (1994) have extended the Peierls framework to 
study thermally activated dislocation emission in certain cases.
(c) Finally, it is possible that significant improvements 
are required in order to provide quantitative theories that 
can predict the intrinsic D/B behavior of a pure material.  
For example, the recent results of 
Xu, Argon and Ortiz (1996b) point to exciting new directions
toward developing quantitative theories of the D/B transition
that include more realistic representation of dislocation nucleation 
processes near a crack tip.

\bigskip

$\dagger$ Present address: Pizzano \& Co. Inc., 92 Montvale Ave., Suite 3750,
Stoneham, MA 02180

$\star$ Corresponding author

\section*{Acknowledgement}

We wish to thank J.R. Rice for useful discussions and Vasily 
Bulatov for a careful 
reading of the manuscript and helpful suggestions.
This work was supported by the Office of Naval Research, 
Contract \# N00014-92-J-1960. 
The calculations were performed at the Pittsburgh Supercomputer Center.

\bigskip

\bigskip

\bigskip

\section*{References}

Argon A.S., Xu G., and Ortiz M. (1996a), in 
{\em Fracture-Instability, Dynamics, Scaling and Ductile/Brittle Behavior},
edited by R. Selinger, J. Mecholksy, A. Carlsson and E. Fuller,
Materials Research Society Symposia Proceedings, vol. 409.

Bachelet G. B., Hamann D. R., and Schl\"{u}ter M., (1982) 
Phys.\ Rev. B {\bf 26}, 4199.

Beltz G.E. and Rice J.R., (1992) Acta Metall. {\bf 40} S321.

Beltz G.E., Rice J.R., Shih C.F., and Xia L. (1996),
Acta Metal. et Mater. (in press).

Denteneer P.J.H., and Soler J.M. (1991a), J. Phys. Cond. Metter {\bf 3}, 8777.  

Denteneer P.J.H., and Soler J.M. (1991b), 
Sold State Comm. {\bf 78}, 857. 

Cheung, K.S. (1990), Ph.D. Thesis, Dept. of Nuclear Eng., 
Massachusetts Institute of Technology, Cabridge, MA.

Ferrante J., and Smith J.R. (1979), Phys. Rev. B {\bf 19}, 3991. 

Griffith A.A., (1920) Phil. Trans. R. Soc. {\bf A 184}, 181. 

Hirth J.P.\  and Lothe J., (1982) 
{\em ``Theory of Dislocations''}, 2nd ed. (Wiley, New York).

Hohenberg P.\ and Kohn W., (1964) Phys. Rev. {\bf 136}, B864.

Juan Y.M, Kaxiras E. and Sun Y., (1995) Phil. Mag. Lett. (to be published).

Kaxiras E.\ and  Duesbery M.S., (1993) Phys. Rev. Lett. {\bf 70}, 3752.

Kohn W.\ and Sham L., (1965) Phys. Rev. {\bf 140}, A1133.

Mehl M.J., and Boyer L.L. (1991), Phys. Rev. B {\bf 43}, 9498. 

Monkhorst H.J.\ and  Pack J.D., (1976) Phys. Rev. B. {\bf 13}, 5188.

Pearson W.B. (1958), {\em A Handbook of Lattice Spacings and Structures of
Metals and Alloys} (Pergamon, New York). 

Peierls R., (1940) Proc. Phys. Soc. London {\bf 52}, 34.

Perdew J.\ and  Zunger A., (1981) Phys. Rev. B {\bf 23}, 5048.

Rice J.R., (1992) J. Mech. Phys. Solids, {\bf 40}, 239.

Rice J.R.\ and Beltz G.E., (1994)  J. Mech. Phys. Solids, {\bf 42}, 333.

Rice J.R., Beltz G.E., and Sun Y., (1992) 
in {\it Topics in Fracture and Fatigue},
edited by A.S. Argon, (Springer, Berlin).

Rice J.R., and Thomson R. (1974), Phil. Mag. {\bf 29}, 73.

Rose J.H., Smith J.R., Guinea F., and Ferrante J. (1984), 
Phys. Rev. B {\bf 29}, 2963. 

Sun Y., Beltz G.E., and Rice J.R., (1993)
Mater. Sci. Engng. {\bf A170} 67.

Sun Y., and Beltz G.E. (1995) J. Mech. Phys. Solids {\bf 42}, 1905. 

Sun Y., Rice J.R., and Truskinovsky L. (1991), in 
{\em High-Temperature Ordered Intermetallic Alloys}, edited by 
L. A. Johnson, D.T. Pope, and J.O. Stiegler, 
Materials Research Society Symposia Proccedings, vol. 213, p. 243.

Wright A., Daw M.S., and Fong C.Y. (1992), Phil. Mag. A {\bf 66}, 387. 

Xu G., Argon A.S., and Ortiz M. (1995), Phil. Mag. A {\bf 72}, 415.

Xu G., Argon A.S., and Ortiz M. (1996b), submitted to Phil. Mag.

Zhou S.J., Carlsson A.E., and Thomson R. (1993), Phys. Rev. B {\bf 47},
7710.

Zhou S.J., Carlsson A.E., and Thomson R. (1994), Phys. Rev. Lett. {\bf 72},
852.

\newpage

\begin{center}
{\bf TABLE I}
\end{center}

\begin{tabular}{||l|c|c|c|c|c||}  \hline
 & $a_0$ (\AA) & B & $C_{44}$ & $C_{11} - C_{12}$ & $\mu$ \\ \hline \hline
Experiment & 4.05 & 76.9 & 28.5 & 46.9 & 26.5 \\ \hline
Present work & 3.95 & 84.8 & 45.5 & 58.8 & 39.1 \\
(at T = 0 K)  & ($-2$\%) & ($+10$\%) & (+60\%) & (+25\%) & (+48\%) \\ \hline
Present work & & & 29.7 & 45.1 & 26.8  \\ 
at $a_0' = 4.05$ \AA          & & & (+4\%) & ($-4$\%) & ($+1$\%) \\ \hline
Mehl and Boyer (1991)  & & & 28.5 & 50.0 & 27.1  \\ 
at $a_0' = 4.05$ \AA       & & & (0\%) & ($+7$\%) & (+2\%) \\ \hline
\end{tabular}

\medskip

TABLE I: Comparison of experimental and theoretical values for 
the lattice constant $a_0$, bulk modulus $B$, elastic 
constants $C_{44}, C_{11}-C_{12}$, and shear modulus $\mu$, evaluated  
at the theoretical zero-temperature equilibrium value 
of the lattice constant (3.95 \AA) and the room-temperature
experimental one (4.05 \AA).  Elastic constants and moduli are given 
in units of GPa and the experimental values are taken from Hirth and Lothe
(1982).  The numbers in parentheses give the percent difference
between theoretical and experimental values.

\bigskip

\bigskip

\begin{center}
{\bf TABLE II} 
\end{center}

\begin{tabular}{||l|c|c|c|c|c|c|c||} \hline
 & $\gamma_{us}^{(u)}$ (J/m$^2$)&$\gamma_{us}^{(r)}$ (J/m$^2$) & $\gamma_s$ 
(J/m$^2$) 
& $b/2(a_0/ \sqrt{6})$ 
& $L/b$ 
& $p = \Delta \theta^{*} / L$ 
& $q = \gamma_{us}^{(u)} / 2\gamma_s$  \\ \hline\hline
DFT/LDA & 0.244 & 0.224 & 1.10 & 0.62 & 0.135 & 0.111 & 0.287  \\
EAM     & 0.092 & 0.079 & 0.57 & 0.50 & 0.140 & 0.0854 & 0.279  \\ \hline
\end{tabular}

\medskip

TABLE II: The values of the unstable stacking energy for 
unrelaxed $\gamma_{us}^{(u)}$ and relaxed $\gamma_{us}^{(r)}$
configurations, the surface energy $\gamma_{s}$, the 
displacement $b/2$ corresponding to the unstable stacking energy along the 
slip route (in units of the 
intrinsic stacking fault slip $a_0/\sqrt{6}$), 
the length scale $L$ for tension, and the scaled 
parameters $p,q$ that determine tension to shear coupling [see text
and Sun, Beltz and Rice (1993)].  
The DFT/LDA values are from the present calculation,
the EAM values from the work of Sun et al. (1993).
The DFT/LDA value of $\gamma_s$ is from Ferrante and Smith (1979).

\bigskip

\bigskip

\begin{center}
{\bf TABLE III}
\end{center}

\begin{tabular}{||c|c|c|c|c||} \hline
Configuration & $\{\vec{a}_1, \vec{a}_2, \vec{a}_3 \}$ & Slip system &
$(\theta, \phi)$ & $G_d / (2 \gamma_s$) \\ \hline \hline
A & \{[$\bar{1}$10],[001],[110]\} & $\frac{1}{6}[121](1\bar{1}1)$ & 
(54.7$^0$,60$^0$) & 1.740 \\
B & \{[001],[$\bar{1}$10],[110]\} & $\frac{1}{6}[\bar{1}12](\bar{1}1\bar{1})$ & 
(35.3$^0$,0$^0$) & 0.968\\
C&\{$[111],[1\bar{1}0],[11\bar{2}]$\}&$\frac{1}{6}[2\bar{1}\bar{1}](11\bar{1})$ & 
(90$^0$,30$^0$) & 0.730\\
D&\{$[1\bar{1}2],[\bar{1}11],[110]$\}&$\frac{1}{6}[\bar{1}12](\bar{1}1\bar{1})$& 
(70.5$^0$,0$^0$) & 0.504\\ \hline
\end{tabular}

\medskip

TABLE III: Ratio of the critical loading for dislocation emission $G_d$
to cleavage energy $(2 \gamma_{s})$, as obtained form the Peierls model
analysis, for various configurations of the slip plane and the 
emited dislocation, characterized by the three vectors  
$ \{\vec{a}_1, \vec{a}_2, \vec{a}_3 \}$ 
and the angles $(\theta, \phi)$ (see text and Fig. 4).  A ratio 
greater than 1 indicates brittle failure, as in configuration A.

\bigskip 

\bigskip 

\bigskip 

FIG. 1:  Illustration of the atomic arrangement in fcc aluminum
and the slab configuration used in the calculations.  Atoms in the three
layers of stacking along the [111] crystallographic direction 
are marked by different symbols.  (a) Top view, (b) side view, indicating
the plane on which the crystal is sheared.

\bigskip

FIG. 2: Convergence tests for the value of the intrinsic stacking 
fault energy of aluminum, 
as a function of plane wave basis size at fixed number of 
sampling points 
(top panel)
and as a function of number of sampling points
in the Brillouin Zone at fixed plane-wave basis size $N_{PW}$    
(bottom panel).  Calculations for slabs of two 
different sizes containing 6 and 9 layers along the [111] direction 
are shown, as well as those from the ANNNI model (see
text for details).

\bigskip

FIG. 3: Generlaized stacking fault energy 
$\gamma_{gsf}$ as a function of displacement along the slip route
in aluminum.  Points are calculated values, the line is a polynomial 
fit.  The maximum in the energy corresponds to the unstable stacking 
energy $\gamma_{us}$, before atomic 
and volume relaxation.  The end point corresponds to the instrinsic 
stacking fault energy $\gamma_{isf}$.  Note that the maximum occurs
slightly to the right of the middle.

\bigskip

FIG. 4: Schematic representation of the geometry for dislocation 
emmision from the crack tip.  The vectors 
$ \{\vec{a}_1, \vec{a}_2, \vec{a}_3 \}$ are along the extension of 
the crack, perpendicular to the crack plane and along the crack line,
respectively.  $\theta$ is the angle between the crack plane and the 
inclined plane on which the dilsocation with Burgers vector $b$ is 
emmited, and $\phi$ is the angle between the emmision direction 
and the Burgers vector.

\end{document}